# Method for precision measurement of the Range - Energy relation for protons in pure hydrogen gas

A.Vorobyev, N. Sagidova

*Petersburg Nuclear Physics Institute*

Contact person: A. Vorobyov
*e*-mail: vorobyov_aa@pnpi.nrcki.ru

December 12, 2019

**Abstract**
The available experimental information on the Range–Energy relation for protons stopped in hydrogen gas is summarized in the SRIM software package. The estimated precision of this data is several percents. Here we describe a possibility to measure this relation with 0.1% precision in the course of an electron–proton elastic scattering experiment to be performed in the 720 MeV electron beam at the Mainz Microtron (MAMI). This experiment, aimed at precision measurement of the proton charge radius, exploits a large hydrogen "active target" to detect the recoiled protons and a forward tracker to detect the scattered electrons. The angles of the scattered electrons are measured with $2 \cdot 10^{-4}$ absolute precision. Also, the electron beam momentum is known at MAMI with $2 \cdot 10^{-4}$ absolute precision. This gives a possibility to determine with 0.1% absolute precision the four-momentum transfer $Q^2$ which, in its turn, is a direct measure of the recoiled proton energy $T_p$: $Q^2 = 2M_p T_p$, where $M_p$ is the proton mass. From this point of view, *this experimental setup can be considered as a source of protons with well defined energies inside an "active target",* which is a specially designed hydrogen high-pressure Time Projection Chamber (TPC). The design of the TPC allows to measure with high precision the energy of the protons corresponding to some selected values of the proton range. In this way, ***the Range–Energy relation can be established for the proton energies from 1 MeV to 9.3 MeV with 0.1% absolute precision,*** the maximal energy being limited by the size of the TPC and by the hydrogen gas pressure. The $H_2$ density in the TPC is controlled on a $3 \cdot 10^{-4}$ precision level in these measurements.

The measured high precision Range –Energy relation for protons in hydrogen gas may have various applications. In particular, it can be used in similar "active target" experiments for precision calibration of the recoiled proton energy scale without requiring precision information on the beam absolute momentum**.** As an example, we consider to apply this method in the *µp* elastic scattering experiment to be carried out in the 100 GeV muon beam at CERN.

**Introduction**

The available information on the Range–Energy relation for the protons stopped in hydrogen gas is summarized in the SRIM and NIST software packages [1]. The stated precision of this data is about 2% for the proton energies $T_p > 1$ MeV and from 2 to 10% for the energies $0.01$ MeV $< T_p < 1$ MeV. Here we describe a possibility to measure this relation with $1 \cdot 10^{-3}$ precision at the proton energies $1$ MeV $\leq T_p \leq 9.3$ MeV in the course of the *ep* elastic scattering experiment [2] to be performed at the Mainz Microtron (MAMI). This experiment, aimed at precision measurement of the proton charge radius, exploits a large hydrogen Time Projection Chamber (TPC) to detect the recoiled protons and a Forward Tracker to detect the scattered electrons. The angles of the scattered electrons will be measured with $2 \cdot 10^{-4}$ absolute precision. Also, the electron beam momentum is known at MAMI with $2 \cdot 10^{-4}$ absolute precision. This gives a possibility to determine with $1 \cdot 10^{-3}$ absolute precision the four-momentum transfer $Q^2$ in the *ep* elastic scattering which, in its turn, is a direct measure of the recoiled proton energy: $Q^2 = 2 M_p T_p$, where $M_p$ is the proton mass. From this point of view, ***this experimental setup can be considered as a source of protons inside the hydrogen TPC with well defined energies***. The TPC was designed at PNPI to measure the energy and the angle of the recoiled protons. In addition, as it is pointed out in this note, this TPC allows also to measure with high precision the mean energy of the protons corresponding to some selected values of the proton range. In this way, ***the Range–Energy relation can be measured with $1 \cdot 10^{-3}$ absolute precision for the proton energies from 1 MeV to 9.3 MeV,*** the maximal proton energy being limited by the TPC size and by the hydrogen gas pressure. The $H_2$ density in the TPC is controlled on a $3 \cdot 10^{-4}$ precision level in these measurements.

Table 1 presents the mean proton projected ranges $<R_p>$ obtained from the SRIM software package for the protons with the energies from 0.01 MeV to 10 MeV stopped in hydrogen gas with the $H_2$ density corresponding to 20 bar pressure at 193 K temperature. For our analysis, we used the following analytical expressions representing the data shown in Table 1:

for $T_p \geq 1$MeV: $\qquad\qquad\qquad <R_p> = 4.82\ T_p^{\ 1.83}$ ; $\qquad\qquad\qquad\qquad\qquad$ (1)

for $0.01$MeV $\leq T_p \leq 1$MeV: $\quad T_p = -0.012 + 0.558 <R_p> - 0.159 <R_p^2> + 0.017 <R_p^3>$, $\quad$ (2)

where the proton range $R_p$ and the energy $T_p$ are expressed in mm and in MeV, respectively.

According to the SRIM package, the longitudinal range straggling is about 4.5% and the lateral straggling is about 1.1 % of the proton range for the proton energies above 1 MeV.

**Table 1** Proton mean projected ranges $<R_p>$ in hydrogen gas for the proton energies $T_p$ from 10 keV to 10 MeV calculated using the SRIM software package. The data corresponds to 20 bar $H_2$ pressure at 293 K temperature, the effects of the gas incompressibility being taken into account

| **Energy $T_p$ MeV** | **0.01** | **0.02** | **0.04** | **0.06** | **0.08** | **0.1** | **0.2** | **0.4** | **0.6** | **0.8** |
|---|---|---|---|---|---|---|---|---|---|---|
| **Range $<R_p>$ mm** | 0.039 | 0.062 | 0.097 | 0.129 | 0.162 | 0.196 | 0.416 | 1.12 | 2.17 | 3.54 |

| **Energy $T_p$ MeV** | **1.0** | **2.0** | **3.0** | **4.0** | **5.0** | **6.0** | **7.0** | **8.0** | **9.0** | **10.0** |
|---|---|---|---|---|---|---|---|---|---|---|
| **Range $<R_p>$ mm** | 5.20 | 17.59 | 36.37 | 61.2 | 91.8 | 127.9 | 169.6 | 216.5 | 268.7 | 326.0 |

**Experimental overview**

The goal of the considered here experiment is to measure the $ep$ differential cross sections in the $Q^2$ region from 0.001 GeV$^2$ to 0.04 GeV$^2$ and to determine the proton charge radius with a sub-percent precision. An innovative method is applied allowing for detection of the recoiled protons and the scattered electrons with high accuracy and resolution.

Figure 1 shows a schematic view of the experimental setup. It consists of a hydrogen Time Projection Chamber (TPC) detecting the recoiled protons and a MWPC based Forward Tracker (FT) detecting the scattered electrons. TPC operates with ultra-clean hydrogen gas at up to 20 bar gas pressure. It allows to measure the ionization produced by the recoiled proton, the recoiled proton angle $\theta_p$, and the longitudinal coordinate of the $ep$ collision vertex $Z_V$. The gas pressure and temperature in TPC are controlled with $1.5 \cdot 10^{-4}$ absolute precision, thus determining the $H_2$ density in TPC with $3 \cdot 10^{-4}$ absolute precision.

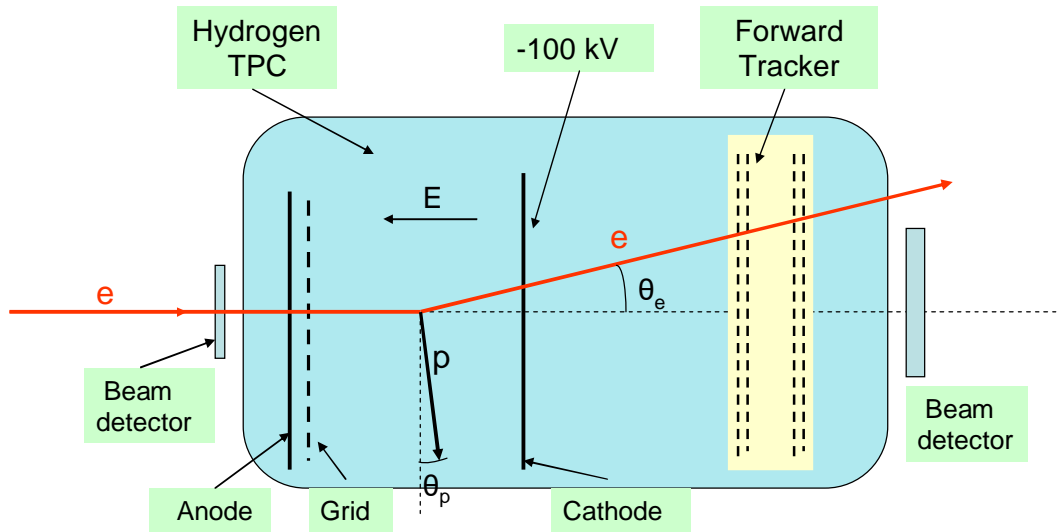

**Fig.1.** The experimental setup for studies of small angle electron-proton elastic scattering. It includes a Hydrogen Time Projection Chamber detecting the recoiled protons and a Forward Tracker (set of cathode strip chambers) detecting the scattered electrons. The absolute energy of the recoiled proton is determined with 0.1% precision from high precision measurements of the electron scattering angle $\theta_e$ and the known absolute energy of the beam electrons. In the context of the present discussion, this setup can be considered as a source of protons with well defined energies inside the hydrogen gas in TPC with precisely determined density. As demonstrated in this note, the TPC allows to establish strict correlation between the proton energy and its mean range in hydrogen for energies from 1 MeV to 9.3 MeV.

The TPC operates in the grid ionization chamber mode (no gas amplification). The drift space is 400.0 mm. The ionization electrons produced by the proton are collected at the TPC anode consisting of a sequence of rings shown in Fig.2. The outer diameter of the anode plane is 600 mm.

The Forward Tracker is designed for high absolute precision in measuring the $X$ and $Y$ coordinates of the electron track relative to the beam line. It consists of two pairs of Cathode Strip Chambers, $X_1/Y_1$ and $X_2/Y_2$. Together with the measured Z coordinate of the vertex, the FT allows to measure the angle $\theta_e$ of the scattered electrons with $2 \cdot 10^{-4}$ absolute precision.

The electron beam enters the TPC through a thin beryllium window. The expected parameters of the beam to be used in this experiment are presented in Table 2.

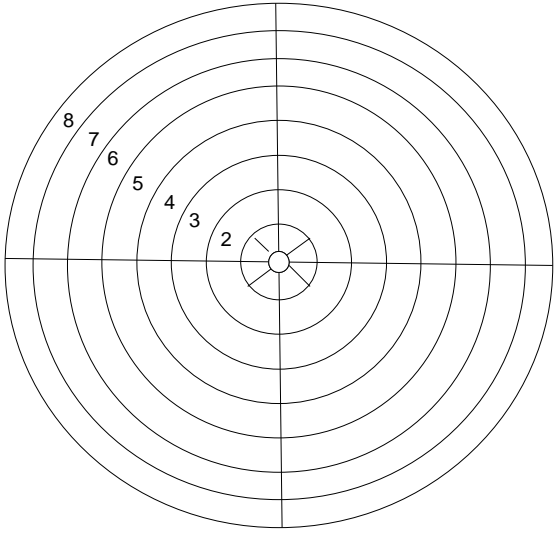

**Fig.2.** TPC anode structure.
The TPC anode plane consists of eight rings around a 1 cm in diameter central pad. The width of each ring, except the outer ring, is 40.0 mm with 0.1 mm gap between the rings. The width of the outer ring is 24 mm. The anode rings are fabricated with 50 μm precision in the radii of the rings. As an example, the radius of the border between Ring 7 and Ring 8 is $\rho_{7/8} = 285.75 \pm 0.05$ mm. All the anode rings, as well as the central pad, have independent readout channels with 20 keV (sigma) energy resolution.

The electron beam enters the TPC through a thin beryllium window. The expected parameters of the beam to be used in this experiment are presented in Table 2.

**Table 2** Parameters of the electron beam from the Mainz Microtron

| Beam energy | 720 MeV |
| --- | --- |
| Beam energy resolution | < 20 keV (1σ) |
| Absolute beam energy precision | ±150 keV (1σ) |
| Beam divergence | ≤ 0.5 mrad (1σ) |
| Beam size | ≤ 0.2 mm (1σ) |
| Beam intensity | $2 \cdot 10^6$ s$^{-1}$ |
| Duty factor | 100 % |

The *ep* elastic scattering differential cross section is determined by the four-momentum transfer $Q^2$. Figure 3 demonstrates the $Q^2$ dependence of the *ep* cross section in the measured $Q^2$ region.

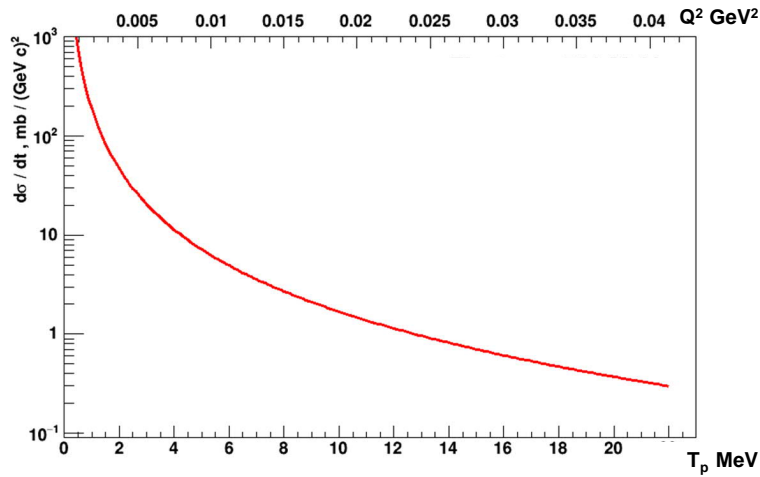

**Fig.3.** Differential cross section of the electron-proton elastic scattering in function of the four-transfer momentum $Q^2 = 2\,M_p T_p$. In the context of this note, it determines the energy spectrum of the protons created in hydrogen gas inside the TPC.

The four-momentum transfer can be determined either by the recoiled proton energy $T_p$ or by the electron scattering angle $\theta_e$:

$$Q^2 = 2M_p T_p; \qquad (3)$$

$$Q^2 = \frac{4\varepsilon_e^2 \sin^2\frac{\theta_e}{2}}{1 + \frac{2\varepsilon_e}{M_p}\sin^2\frac{\theta_e}{2}}, \qquad (4)$$

where $\varepsilon_e$ is the total energy of the beam electron at the moment of the *ep* collision.

In this experiment, the value of $Q^2$ is determined from measurements of the recoiled proton energy $T_p$. Note, however, that what is measured with the TPC is not directly the proton energy $T_p$ but the ionization charge produced by the proton. Therefore, some calibration procedure is required which could relate the observed signals from the TPC with $T_p$. In this experiment, such calibration is foreseen using the absolute $Q^2$ values determined from the measured $\theta_e$ distributions and from the known absolute value of the beam energy $\varepsilon_e$. The stated absolute precision of the $T_p$ scale calibration is $8 \cdot 10^{-4}$. *So, in the context of the presented here note, this experimental setup can be considered as a source of protons inside the hydrogen TPC with well defined energies.* The energy distribution of the protons follows the *ep* differential cross section (Fig. 3). The protons are emitted nearly perpendicular to the beam axis and, as it will be shown below, their mean projected range can be measured using the TPC signals. This gives a unique possibility for high precision determination of the Range – Energy relation for protons in hydrogen gas in a certain $T_p$ region.

**Method for measurements of the mean proton ranges in TPC**

The protons are emitted from the TPC central zone (1mm radius around the beam line) at the angle $\theta_p$ given by the following expression:

$$\sin\theta_p = [(\varepsilon_e + M_p)/P_e] \cdot T_p/P_p, \qquad (5)$$

where $P_p$ are $P_e$ are the momenta of the recoiled proton and the beam electron, respectively, at the point of the *ep* collision,. The TPC anode sees the projection of the proton range:

$$R_p^* = R_p \cos\theta_p. \qquad (6)$$

For the Mainz TPC experiment ($\varepsilon_e = 720$ MeV) this gives:

$$R_p^* = R_p (1 - 0.0014\, T_p\,(\text{MeV})). \qquad (7)$$

Now we can modify Eq. (1) so that it would relate the proton energy $T_p$ with the projection of the proton range measured in TPC:

$$\langle R_p^* \rangle = 4.82\, T_p^{1.83} (1 - 0.0014\, T_p). \qquad (8)$$

The ionization electrons produced by the proton are collected at the TPC anode consisting of a sequence of rings shown in Fig. 2. The width of each ring, except the outer ring, is 40.0 mm with 0.1 mm gap between the rings. The width of the outer ring is 24 mm. The anode rings are fabricated with 50 µm precision in the radii of the rings.

The proton ranges have rather large straggling. According to the SRIM data, the longitudinal straggling is $\Delta R_p(\text{sigma}) \approx 0.045 \langle R_p \rangle$ in the considered proton energy region from 1 MeV to 10 MeV. This sets the precision limit in the proton range measurements on the event-by-event basis.

As to the proton Range-Energy relation, it is formulated for the mean proton ranges corresponding to the fixed proton energies, and this is what can be measured with the TPC. In fact, this is not even a separate measurement, it is just a special analysis of the main experimental data.

The idea of the method is to select from the available experimental data set a sample of protons in a $\Delta T_p$ bin around some value of $T_p$ with the protons stopped in average exactly against the border between the neighbour rings (ring $J$ and ring $K$). The radius of the border $\rho_{j/k}$ being known with 0.02% precision, this determines the mean proton range (and thus the Range-Energy relation) with high precision.

The procedure of such selection might be as follows. As a first step, one can select the protons with the mean range $<R_p^*>$ slightly below $\rho_{j/k}$. The corresponding value of $T_p$ can be calculated with the help of Eq.(8). The signal on the outer ring $K$ appears when the proton range exceeds the middle of the gap between the rings, that is when $R_p^* > \rho_{j/k}$. The ratio of the number of penetrating to the outer ring protons $N_>$ to the total number of the selected protons $N_{tot}$ is a measure of the distance between $<R_p^*>$ and the rings border radius, while **the value $N_>/N_{tot} = 0.50$ corresponds to the equality $<R_p^*> = \rho_{j/k}$.** The optimal proton energy $T_p^*$ corresponding to $N_>/N_{tot} = 0.50$ can be found from scanning the region around $\rho_{j/k}$ with variation of the selected proton energy $T_p$, thus determining the Energy-Range relation in this energy point. Simultaneously, similar measurements can be done using the proton stops in the border regions between all TPC rings. The measurements can be performed at various gas pressures from 1 bar up to 20 bar allowing to measure the proton Range-Energy relation in hydrogen gas for proton energies from 1 MeV to 9.3 MeV. Below we present the results of a MC simulation of such analysis demonstrating the precision of this method.

**Monte Carlo simulation of the proton range measurements in TPC**

For this analysis, we generated samples of the protons with fixed (variable) energy $T_p$. The proton range distribution $F_{Tp}(R_p^*)$ for such samples was taken as a Gaussian with the mean value $<R_p^*>$ and the dispersion $\sigma_{Rp^*} = 0.05<Rp^*>$, to take into account the longitudinal straggling of the proton ranges, with $R_p^*$ given by Eq.(7). We have considered the case of the protons stopped against the region between Ring7 and Ring8 at 20 bar $H_2$ pressure. The radius of the border between these rings is $\rho_{7/8} = 285.75 \pm 0.05$ mm.

Seven $F_{Tp}(R_p^*)$ distributions were generated for the proton energies from $T_p = 9.20$ MeV to $T_p = 9.50$ MeV with $<R_p^*>$ calculated using Eq.(8). In each distribution, 10000 protons ($N_{tot}$) were generated, and the number $N_>$ of the protons penetrating into the zone of the outer ring (that is with $R_p^* > 285.75$ mm) was counted. The obtained ratio $N_>/N_{tot}$ is presented in Fig.4. One can see that a linear fit to the $T_p$ dependence of the $N_>/N_{tot}$ ratio is satisfactory in this $T_p$ region. From this fit, one can find the proton energy $T_p^*$ corresponding to the ratio $N_>/N_{tot} = 0.50$ (that is to $<R_p^*> = 285.75$ mm), as well as the precision of thus determined $T_p^*$ value.

In order to study better the precision of the $T_p^*$ value extracted with this method, the described above procedure (generation of seven $F_{Tp}(R_p^*)$ distributions and finding $T_p^*$ from the fit shown in Fig.4 ) was repeated 1000 times. The obtained $T_p^*$ distribution is presented in Fig. 5. As one can see from this figure, the statistical error in the extracted $T_p^*$ value is quite low (RMS = $1.3 \cdot 10^{-4}$ $T_p^*$) But it was obtained with rather high generated statistics (10000 protons in each $F_{Tp}(R_p^*)$ distribution). Figure 6 presents similar $T_p^*$ distribution obtained with lower statistics (1000 events in each $F_{Tp}(R_p^*)$ distribution). In this case, the precision of the extracted $T_p^*$ is $4.2 \cdot 10^{-4}$ which is still sufficient for determination of the Range – Energy relation with 0.1% precision.

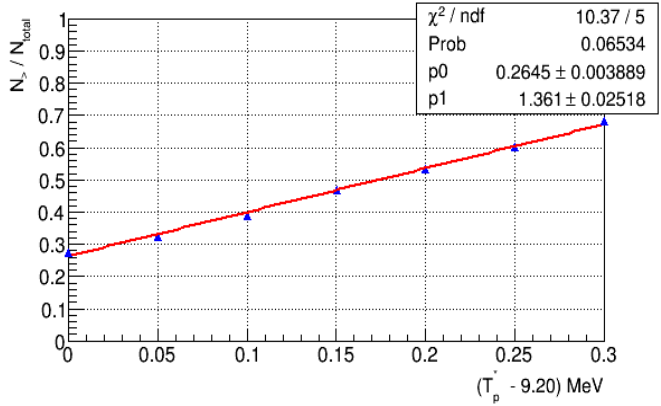

**Fig. 4.** The ratio $N_>/N_{tot}$ of the number of penetrating to the outer ring protons to the total number of the generated protons measured for seven $F_{Tp}(R_p^*)$ distributions in the $T_p$ region from 9.150 MeV to 9.450 MeV with 10000 generated events in each distribution. The linear fit to this data determines the proton energy $T_p^*$ corresponding to the ratio $N_>/N_{tot} = 0.50$ where $<R_p^*> = \rho_{7/8}$

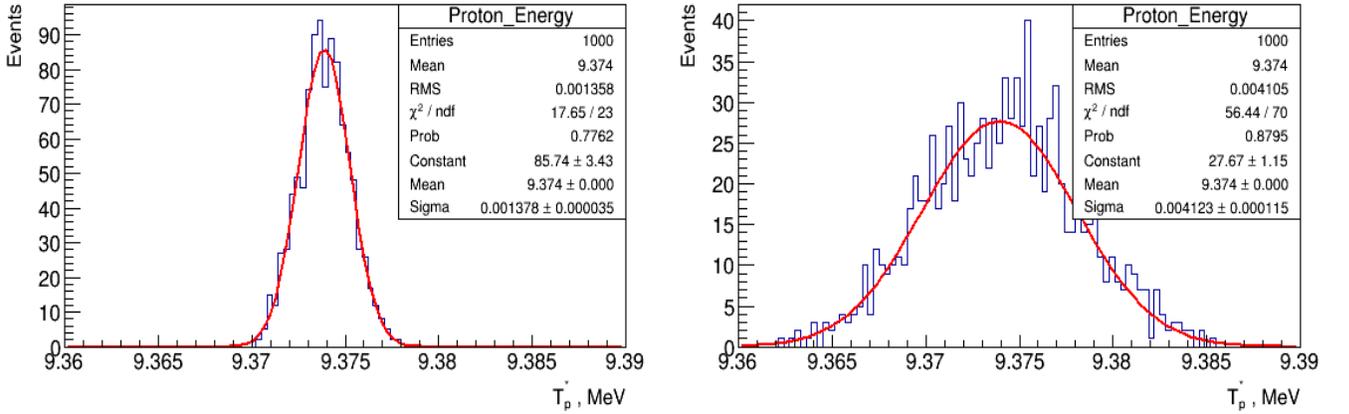

**Fig.5**. Distribution of the recoiled proton energy $T_p^*$ corresponding to the ratio $N_>/N_{tot} = 0.50$ obtained with the described in the text procedure repeated 1000 times. The $F_{Tp}(R_p)$ distributions used in this procedure contained 10000 generated events in each distribution (Left panel) and 1000 events (Right panel).

In the real experiment, the $F_{Tp}(R_p^*)$ distribution will be replaced by the *ep* elastic scattering events selected in a $\Delta T_p$ bin at some $T_p$ energy (see Fig.3), and the ratio $N_>/N_{tot}$ will be measured in the scanning procedure as described above. Note that the number of events in a 50 keV $\Delta T_p$ bin is about 20000 at $T_p \approx 10$ MeV for the full statistics expected in the described here experiment. That means that the Range–Energy relation in this energy point can be measured with 0.1% precision already with ~ 5% of the total statistics.

We have considered above the case when the selected protons were stopped close to the border between Ring 7 and Ring 8. Similar procedure can be applied also for the other neighbour rings. In this way, the measurements of the Range–Energy relation can be performed simultaneously in several points in the $T_p$ region from 5 MeV to 9.3 MeV at 20 bar hydrogen pressure, and expanded down to $T_p = 1$ MeV at 1 bar pressure.

The signals from the outer ring will be detected using the readout electronics with ~ 100 keV threshold (20 keV noise). This might reduce the measured number $N_>$ because of the losses of the low amplitude signals below the threshold, thus leading to some shift of the measured $T_p^*$ to lower values. Figure 7 (Left panel) shows the simulated range distribution of the protons penetrated into the zone of Ring 8 calculated for $F_{Tp}(R_p^*)$ with $<R_p^*> = \rho_{7/8} = 285.75$ mm and $\sigma(R_p^*) = 0.05<R_p^*>$. Also shown is the amplitude distribution of the corresponding TPC signals at Ring 8. In these calculations, the range-to-energy transfer was performed according to the SRIM tables.

As it follows from Fig.7 (Right panel), the number of the signals below the 100 keV threshold is ~ 1% of the total number of the signals detected at Ring 8. In principle, such losses may shift the value of $T_p^*$ by 0.15%. However, the number of the lost events could be restored with ~10% precision by extrapolating the measured amplitude distribution of the signals to the region below 100 KeV, thus making this effect negligible.

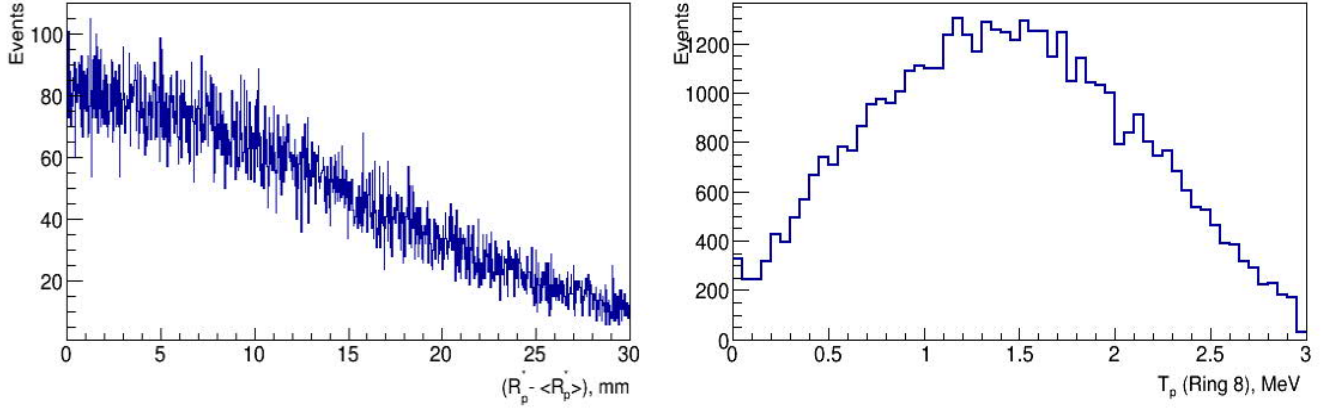

**Fig.7. Left panel**: Range distribution of the protons penetrated into the zone of Ring 8, calculated using a Gaussian proton range distribution $F_{Tp}(R_p^*)$ with $<R_p^*> = \rho_{7/8} = 285.75$ mm and $\sigma(R_p^*)=0.05<R_p^*>$. The number of generated events in $F_{Tp}(R_p^*)$ is $N_{tot} = 100\,000$. **Right panel**: Amplitude distribution of the corresponding signals on Ring 8. The number of the signals below the 100 KeV threshold is ~ 0.5% of the total number $N_{tot}$. It can be restored by extrapolation of the amplitude distribution to the region below 100 KeV.

**Discussion**

In this note, we have pointed out a possibility to measure the proton Range-Energy relation in hydrogen gas in the course of an experiment at MAMI aimed at high precision measurement of the proton charge radius in the *ep* elastic scattering. The basic element of this experiment is a large hydrogen filled TPC detecting emitted in the *ep* collisions protons. The energy of the protons is determined with high absolute precision via the angle and the momentum of the scattered electrons, due to precisely measured absolute energy of the beam electrons at MAMI ($2 \cdot 10^{-4}$ absolute precision).

In the context of this Note, this experimental setup can be considered as a source of protons emitted from the central zone of the TPC gas volume with continuous spectrum of well defined energies. Using the described in this Note method of analysis of the TPC data, one can establish with 0.1% precision the Range-Energy relation for the protons with energies from 1 MeV to 9.3 MeV. This precision is an order of magnitude higher than the cited precision of the SRIM data for the proton energies above 1MeV. It should be noted that the SRIM software package above 1MeV is based mainly on the theoretical approach. The new experimental data may be used to validate this approach and to improve the precision of the proton ranges calculated with the SRIM software package.

One remark on definition of the proton range. The requirement $N_>/N_{tot} = 0.50$ in the described above method of measuring the proton Range-Energy ratio corresponds, strictly speaking, to finding the median of the range distribution at some fixed proton energy. Therefore, the used in this article notation $<R_p>$ is meaning, in fact, the median of the range distribution. In the case of the Gaussian (or any symmetric) distribution, the median coincides with the mean value.

The proton energy $T_p$ plays an exceptional role in studies of the lepton-proton (*ep* or *μp*) elastic scattering as it is a direct measure of the four-transfer momentum: $Q^2 = 2M_p T_p$. In the low $Q^2$ region (which is the field-of-interest in measurements of the proton charge radius), the differential cross sections of the lepton-proton elastic scattering at relatively high beam energies are practically independent on this energy, they are fully determined by $Q^2$, that is by the recoiled proton energy $T_p$. This offers an attractive possibility to perform measurements of the differential cross sections without requiring precision information on the absolute momentum of the scattered beam particle. However, this information remains mandatory for the absolute calibration of the $T_p$ scale, if the $T_p - \theta_e$ calibration method is used for this purpose. As an example, such calibration of the $T_p$ scale in the considered above TPC experiment at Mainz is possible only due to precisely measured energy of the beam electrons at MAMI.

Recently, a new TPC experiment was proposed for the proton radius measurements via studies of the *μp* elastic scattering in the 100 GeV muon beam at CERN. In this experiment, needed for the $T_p - \theta_\mu$ calibration method, determination of the absolute momentum of the scattered muons on a ~ 0.1% precision level is a rather complicated task. On the other hand, when the proton Range–Energy relation measured to 0.1% precision will be available, it could be used for the $T_p$ scale absolute calibration using the described above procedure ( the $N_>/N_{tot} = 0.50$ method). In this way, the $T_p$ scale will be calibrated to ~ 0.1% absolute precision without applying any information on the beam momentum. Note also that the $T_p - R_p$ calibration method will not require any additional equipment or measurements, the calibration will be performed on a bases of analysis of the main data set.

Apart from the absolute calibration of the $Q^2$ scale in the TPC experiment at CERN, needed for the measurement of the proton charge radius, the $T_p - R_p$ calibration method guaranties that the $Q^2$ scales used in both the *ep* and the *μp* experiments are closely identical, that provides a unique possibility for these experiments to ***study the muon-electron universality in lepton scattering in the most precise way***. In this context, it is also important that the design parameters of the TPCs are practically identical in both experiments which makes the comparison of the measured differential cross sections even more valuable.

**References**

1. SRIM version -2013 J.P Biersack and J.F.Ziegler 2008 ;
2. V.Vorobyev. Precision measurement of the proton charge radius in electron proton scattering. Report at International Conference "Hadron Structure and QCD (HSQCD2018)". Gatchina, Russia, 6-10 August, 2018. Physics of Particles and Nuclei Letters, 16(5), 524-529.